\documentclass[twocolumn,superscriptaddress,amsfont,amssymb,amsmath, showpacs,balancelastpage, nofootinbib]{revtex4-1}
\usepackage{graphicx,longtable,mathrsfs,color,array}
\usepackage[hidelinks]{hyperref}
\usepackage[usenames,dvipsnames]{xcolor} 
\usepackage{amssymb,amsmath,mathtools,mathrsfs,slashed} 
\usepackage{epsfig,subfigure,placeins,float} 
\usepackage{booktabs,longtable,ctable,multirow} 
\usepackage{exscale,relsize} 
\usepackage[normalem]{ulem} 
\usepackage{enumerate}
\usepackage{times, mathptmx} 
\usepackage{comment}
\usepackage{color}
\usepackage[mathscr]{euscript}
\allowdisplaybreaks[1]

\begin{document}
\title{Standard Sirens as a novel probe of dark energy} 
\author{William J. Wolf}
\email{wwolf@uchicago.edu}
\affiliation{Kavli Institute for Cosmological Physics, The University of Chicago, Chicago, IL 60637, USA}
\author{Macarena Lagos}
\email{mlagos@kicp.uchicago.edu}
\affiliation{Kavli Institute for Cosmological Physics, The University of Chicago, Chicago, IL 60637, USA}
\date{Received \today; published -- 00, 0000}

\begin{abstract}
Cosmological models with a dynamical dark energy field typically lead to a modified propagation of gravitational waves via an effectively time-varying gravitational coupling $G(t)$. The local variation of this coupling between the time of emission and detection can be probed with standard sirens. Here we discuss the role that Lunar Laser Ranging (LLR) and binary pulsar constraints play in the prospects of constraining $G(t)$ with standard sirens. In particular, we argue that LLR constrains the matter-matter gravitational coupling $G_N(t)$, whereas binary pulsars and standard sirens constrain the quadratic kinetic gravity self-interaction $G_{gw}(t)$. Generically, these two couplings could be different in alternative cosmological models, in which case LLR constraints are irrelevant for standard sirens. We use the Hulse-Taylor pulsar data and show that observations are highly insensitive to time variations of $G_{gw}(t)$ yet highly sensitive to $G_N(t)$. We thus conclude that future gravitational waves data will become the best probe to test $G_{gw}(t)$, and will hence provide novel constraints on dynamical dark energy models.
\end{abstract}

\date{\today}
\maketitle


\textit{Introduction.---} 
Gravitational waves (GW) are a long-standing prediction of Einstein's theory of general relativity (GR) and their recent direct detection by the LIGO/Virgo collaboration \cite{LIGOScientific:2018mvr} has confirmed many aspects of GR \cite{TheLIGOScientific:2016src, Abbott:2018lct, LIGOScientific:2019fpa} while future tests will offer exciting opportunities to probe the Universe and its constituents even further.

In particular, the unknown nature of dark energy -- the component driving the observed late-time accelerated expansion of the Universe \cite{Riess:1998cb} --
has motivated a number of alternative models that promote the cosmological constant of the standard $\Lambda$CDM model to be a dynamical dark energy field (see e.g.~\cite{Nojiri:2010wj, Clifton:2011jh, Joyce:2014kja, Koyama:2015vza, Joyce:2016vqv, Nojiri:2017ncd,  Frusciante:2019xia}).
GW will provide important constraints for cosmology via standard sirens (multi-messenger detections of GW and electromagentic signals \cite{0004-637X-629-1-15}), as they can test the properties of these dynamical dark energy fields as well as provide independent constraints on cosmological parameters such as the current expansion rate of the universe, $H_0$, whose current observational constraints exhibit a 4-6$\sigma$ tension (see review in \cite{Verde:2019ivm}). 
In the upcoming years, detectors such as KAGRA \cite{Akutsu:2018axf}, LIGO-India \cite{Unnikrishnan:2013qwa}, the ET \cite{Sathyaprakash:2012jk} and LISA \cite{Klein:2015hvg} (and possible proposed detectors such as DECIGO \cite{Seto:2001qf}, Voyager \cite{LIGO}, and CE \cite{Reitze:2019iox}) will come online to measure GW with more precision and over cosmological distances.

One notable non-trivial signature that can be probed with standard sirens is the possibility of a time-varying gravitational coupling $G(t)$, which typically arises in dynamical dark energy models. In this case, the quadratic action for GW that propagate at the speed of light on a cosmological background is typically given by:
\begin{equation}\label{GWAction}
    S=\int dtd^3x\, \frac{a^3(t)}{16\pi G(t)c^2}\left[ \dot{h}_A^2-\frac{c^2}{a^2}(\vec{\nabla} h_A)^2\right],
\end{equation}
where overdots denote time derivatives, $a$ is the scale factor with rate $H=\dot{a}/a$, $c$ is the constant speed of light, and $h_A$ describes the amplitude of GW for its two possible polarizations $A=+,\times$. Here, $G(t)$ quantifies the kinetic quadratic self-interaction coupling of the spacetime metric, and depends on the cosmological evolution of the dark energy field. Constraining the time evolution of $G(t)$ is important because it can be highly degenerate with $H_0$ \cite{Lagos:2019kds} and therefore its presence can bias $H_0$ measurements. This happens because it leads to a friction term in the propagation of GW. In particular, the equation of motion for GW from eq.~(\ref{GWAction}) is given by:
\begin{equation}
    \ddot{h}_A + [3+\alpha_M]H\dot{h}_A +(k/a)^2h_A=0,
\end{equation}
where $\alpha_M=-(1/H)(\dot{G}/G)$ and $k$ is the wavenumber in Fourier space. This new friction term vanishes in $\Lambda$CDM, and therefore any deviation from zero would imply a non-trivial nature of dark energy.

A particular way to test $\alpha_M$ is with multi-messenger detections, of both electromagnetic (EM) and GW signals, through observations of the electromagnetic luminosity distance $d^{em}_{L}$ and the so-called GW luminosity distance $d^{gw}_{L}$--which probes the decay of the GW amplitude as it travels over cosmological distances. Specifically, these distances are related by \cite{Deffayet:2007kf, Saltas:2014dha, Lombriser:2015sxa, Arai:2017hxj, Amendola:2017ovw, Belgacem:2017ihm}: 
\begin{equation}
d^{gw}_{L}(z)/d^{em}_{L}(z)  =\sqrt{G(z)/G(0)},
\end{equation}
and hence probing the observables in the LHS of this equation would provide constraints on the time evolution of the self-interaction gravitational coupling. 
Indeed, recent forecasts show that standard sirens could impose bounds that range between $|\dot{G}/G|\lesssim  \mathcal{O}(1)H_0$ for LIGO \cite{Lagos:2019kds}, and $|\dot{G}/G|\lesssim  \mathcal{O}(10^{-2})H_0$ for  LISA \cite{Belgacem:2019pkk} (and similar bounds for other detectors such as ET, DECIGO, Voyager, and CE \cite{Nishizawa:2019rra, DAgostino:2019hvh, Bonilla:2019mbm}).

In \cite{Amendola:2017ovw, Dalang:2019fma} it was argued and clarified that what standard sirens probe is the difference in the {\it local} value of $G$ at the moment of emission and detection (as opposed to its cosmological value). Because of this, other small-scale constraints on time varying gravitational couplings could already give insights on the outlook of standard sirens for testing $G$. On this regard, in \cite{Tsujikawa:2019pih, Dalang:2019fma} it was shown that because Lunar Laser Ranging (LLR) data constrains $|\dot{G}/G|\lesssim \mathcal{O}(10^{-3})H_0$ \cite{Hofmann:2018myc}, then future standard sirens would not have the sensitivity to further probe the effects coming from a varying $G$. 

In this Letter, we analyze a subtle but crucial aspect to take into consideration when combining standard sirens with LLR measurements. We argue that local measurement constraints from LLR probe the matter-matter interaction $G_N$, whereas standard sirens probe the quadratic kinetic metric self-interaction $G_{gw}$ (a similar distinction between couplings has been made in \cite{Tahura:2018zuq, Carson:2019kkh} in the context of waveforms). In GR, these two couplings are constants and equal, given by Newton's constant. However, when dark energy is a dynamical field and allowed to have non-trivial interactions with the metric, these two couplings may very well differ in magnitude and evolution. In fact, we will show that very simple models of dark energy considered in the literature already exhibit $G_N\not=G_{gw}$. We therefore conclude that LLR constraints are not generically applicable to standard sirens measurements.

Motivated by the previous argument, in this Letter we also explore whether other local observations may provide constraints on $G_{gw}$, and consequently affect the outlook of standard sirens even if $G_{N}\not=G_{gw}$. In particular, we focus on  binary pulsars as they indirectly test gravitational waves and hence should be sensitive to time variations of $G_{gw}$. By observing the time variation of the period of the binary, one can probe the loss of energy by GW and a potential time varying coupling. It has already been found from various binary pulsars \cite{1975ApJ...195L..51H, Splaver:2004du, Nice:2005fi, Verbiest:2008gy, Deller:2008jx, 2009MNRAS.400..805L, Manchester:2015mda} that $|\dot{G}/G|\lesssim \mathcal{O}(10^{-2}) H_0$. However, these analyses were all done assuming that $G_N=G_{gw}$. In this Letter we revisit this constraint assuming that $G_N\not=G_{gw}$, and use the Hulse-Taylor pulsar as representative to get general estimates on constraints of the time evolution of $G_{gw}$ and $G_N$. We find that while technically the time variation of both $G_N$ and $G_{gw}$ affect the period of the orbit, the effects coming from $G_{gw}$ are suppressed. As a result, binary pulsar data does have the sensitivity to observe changes in $G_N$ of order $H_0$, but it does not have the sensitivity to observe $|\dot{G}_{gw}/G_{gw}|\sim \mathcal{O}(1)H_0$. We thus conclude that for models that evade LLR ($G_N\not=G_{gw}$), then standard sirens will provide the most promising probe to test the time dependence of $G_{gw}$ in the future.

\textit{Distinguishing Interactions.---} 
In gravity models with minimally coupled matter (such as the $\Lambda$CDM model) the only relevant gravitational coupling is Newton's constant. If non-minimal couplings are allowed, such as in typical dynamical dark energy scalar models, the equivalence principle is violated and we then must clearly specify the fields and bodies interacting in order to determine the relevant coupling parameter. 
In particular, we will be distinguishing two physical couplings: $G_N$ describing the local interaction of gravity between two massive bodies, and $G_{gw}$ the local kinetic quadratic self-interactions of the gravitational field. In this Letter, we will remain agnostic about the underlying cosmological model and the mechanisms for generating different couplings $G_N$ and $G_{gw}$. 
Nevertheless, as an example, we mention that simple scalar-tensor theories can lead to $G_N\not=G_{gw}$. In particular, theories equipped with the chameleon mechanism have an action given by (see e.g.~reviews \cite{Khoury:2003rn, Burrage:2017qrf}):
\begin{equation}\label{ChameleonAction}
S=\int d^4x\; \sqrt{-g}\left[\frac{c^3A^{-2}(\varphi)}{16\pi G_p}R- \frac{1}{2}k(\varphi)\partial^\mu \varphi \partial_\mu\varphi -\frac{V(\varphi)}{A^4(\varphi)}\right],
\end{equation}
where $G_p$ is the usual Newton's constant, $R$ is the Ricci scalar, $\varphi$ is the dark energy field, and $A$, $k$ and $V$ are appropriate functions of $\varphi$. Here it is assumed that additional matter components are minimally coupled to the metric $g_{\mu\nu}$. On the one hand, for typical choices of these functions $A$, $k$, and $V$, it has previously been found that the local matter-matter coupling between two spherically symmetric bodies $1$ and $2$ is given by $G_N\approx G_p(1+2\beta^2Q_1Q_2)$ \cite{Mota:2006fz, Sakstein:2017pqi}, where $\beta$ is a dimensionless model parameter (usually assumed to be order 1), $Q_{1,2}\ll 1$ are dimensionless quantities depending on the bodies $1$ and $2$ and are much smaller than one when screening is effective (which is in turn determined by both the density of the bodies and the galactic density they are embedded in).

On the other hand, the quadratic kinetic self-interaction of the metric is obtained by considering small perturbations $h_{\mu\nu}$ of the metric around a spherical background, and identifying $G_{gw}$ with the coefficient multiplying the quadratic terms in $\dot{h}_{\mu\nu}$ when expanding the Action (\ref{ChameleonAction}) (in a similar manner to eq.~(\ref{GWAction})). For the Chameleon model, $G_{gw}$ will be given by the conformal coupling and therefore, around a dense body embedded in a galactic background, we will have $G_{gw}=G_pA^{2}(\varphi)\approx G_p(1 + 2\beta\varphi_\text{gal}\sqrt{G_p}/c^2)$, where $(\varphi_\text{gal}\sqrt{G_p}/c^2)\ll 1$ is the value of the scalar field in the galactic environment in Planck units.
Due to the similarities between Chameleon and Symmetron screening mechanisms, Symmetron theories will also have $G_N\not=G_{gw}$. This example illustrates how the simplest dynamical dark energy models already exhibit different couplings. These calculations have been performed on a static universe but can be generalized to include a cosmological evolution where the environmental galactic density will depend on time. This will induce a time evolution on both the strength and range of effectiveness of screening, and since the couplings themselves are different, we expect their time evolution to be different as well. Note that for Chameleon and Symmetron models we expect the time evolution of both couplings to be much smaller than one Hubble time. However, models with other screening mechanisms may not suppress the time evolution of both couplings \cite{Kimura:2011dc}, so in this Letter we remain agnostic and consider the possible time evolution of these two couplings to be arbitrary a priori.


Next, we focus on observational constraints for the time variation of these couplings. LLR data already place tight constraints on the possible time evolution of the gravitational coupling, setting $-0.07\times 10^{-3}(0.7/h)H_0<\dot{G}_N/G_N < 2.05\times 10^{-3}(0.7/h)H_0$ \cite{Hofmann:2018myc}, where $H_0=100 h$ km/s/Mpc. This constraint is found by probing the gravitational force between the Earth and the Moon, and therefore probes $G_N$. Another important constraint comes from the Hulse-Taylor binary pulsar $-0.8\times 10^{-2}(0.7/h)H_0<\dot{G}_N/G_N < 12.8\times 10^{-2}(0.7/h)H_0$ \cite{BisnovatyiKogan:2005ap}, which was obtained assuming $G_N=G_{gw}$. 

\textit{Varying $G_{gw}$ and $G_N$ on Binary Pulsar.---}
Next, we study the emission of GW from a binary of two point masses and its effect on the binary's period decay, assuming $G_N\not=G_{gw}$. In this analysis, we will assume that the only non-trivial effect in the motion of the binaries is due to the presence of a time-varying $G_N$.
Similarly, we also assume that the local equation of motion for gravitational waves also holds modulo the time-varying coupling, and hence we describe models that still have second-order derivative equations of motion (see e.g.~\cite{Banerjee:2016nnn} for a similar analysis in a model with higher derivatives). We start by considering the linearized Einstein equations in the Lorentz gauge with a $G_{gw}(t)$, valid in the weak field regime:
\begin{equation}\label{EqhT}
    \Box{h}_{\mu\nu}=\frac{16 \pi G_{gw}}{c^4} T_{\mu\nu}
\end{equation}
where $\Box=\partial_\mu\partial^\mu$, and $T_{\mu\nu}$ is the stress-energy momentum tensor of matter, which we consider to be given by a binary system of point masses, which will in turn depend on $G_N$. Thus, gravitational radiation will depend on both $G_{gw}$ and $G_N$, as we later show in eq.~(\ref{power}).
We also note that $G_{gw}$ comes from the linearized Einstein equations and it corresponds to the quadratic kinetic metric self-interaction. Indeed, one can check for Chameleon models that eq.\ (\ref{EqhT}) holds with $G_{gw}$ determined by the conformal factor.
After choosing the transverse-traceless gauge (where $h_{\mu\nu}$ is traceless and with spatial non-zero components $h^{TT}_{ij}$ only) and performing a non-relativistic expansion of $T_{ij}$, one finds at leading order 
\cite{Maggiore:1900zz}:
\begin{align}\label{Solh}
    &h^{TT}_{ij}=\frac{2G_{gw}}{c^4r}\Ddot{Q}^{TT}_{ij},
\end{align}
where $r$ is the comoving distance between the source and a far away observer. Also, $\ddot{Q}_{ij}$ is the second time derivative of the traceless mass quadrupole moment, which is defined as $Q_{ij}=\int d^3x\,\rho(t,\vec{x})(x^ix^j-\delta^{ij}r^2/3)$ with $\rho=T^{00}/c^2$. 
Note that eq.~(\ref{Solh}) is expressed in terms of $Q_{ij}^{TT}$, whose only non-zero components are those transverse to the propagation direction of the wave. Next, we consider the radial power radiated over a solid angle by these GW, which is generically given by \cite{Maggiore:1900zz}:
\begin{equation} \label{oldP}
    \frac{dP_{gw}}{d\Omega} = \frac{r^2c^3}{32 \pi G_{gw}} \langle \dot{h}^{TT}_{ij} \dot{h}^{TT}_{ij} \rangle .
\end{equation}
Here, brackets denote a temporal average on the oscillations of the GW. 
Since $G_{gw}$ varies in time, when calculating $\dot{h}^{TT}_{ij}$ in terms of $Q_{ij}$ we obtain the following power:
\begin{align}\label{Pgw1}
    P_{gw} &= \frac{G_{gw}}{5c^5}\left[  \langle \dddot{Q}_{ij} \dddot{Q}^{ij} \rangle + \frac{2\dot{G}_{gw}}{G_{gw}} \langle \Ddot{Q}_{ij} \dddot{Q}^{ij}\rangle + \frac{\dot{G}_{gw}^2}{G_{gw}^2} \langle \Ddot{Q}_{ij} \Ddot{Q}^{ij} \rangle  \right].
\end{align}
Here we see that by allowing the kinetic quadratic self-interaction of gravity to vary, two new terms beyond GR appear in the quadrupole power formula. 

Next, we calculate each one of the terms in eq.~(\ref{Pgw1}) explicitly. 
We start by considering an elliptical orbit akin to the Hulse-Taylor binary orbiting in the $(x, y)$ plane with polar coordinates given by $(r, \psi)$. The masses of the two objects are denoted $m_1$ and $m_2$.
We write the mass moment $M_{ij}$ of this system in terms of its reduced mass $\mu=m_1m_2/(m_1+m_2)$, the total mass $m=m_1+m_2$, semi-major axis $a$, and orbital eccentricity $e$ \cite{Maggiore:1900zz}:
\begin{equation}\label{Mij}
M_{ij} = 
 \frac{\mu a^2(1-e^2)^2}{(1+e \cos(\psi)^2)}
\begin{pmatrix}
\cos^2(\psi)& \cos(\psi)\sin(\psi)\\
\cos(\psi)\sin(\psi)& \sin^2(\psi)\\
\end{pmatrix}.
\end{equation}
This mass moment is related to the traceless mass quadrupole by $Q_{ij} = M_{ij} - \delta_{ij}M_{kk}/2$.
Furthermore, the angular velocity $\Omega \equiv \dot{\psi}$ is also related to these orbital parameters through the conservation of angular momentum:
\begin{equation}\label{Omega}
    \Omega=\left(\frac{G_N m}{a^3}\right)^{1/2}(1-e^2)^{-3/2} (1+e\cos(\psi))^2,
\end{equation}
which depends on the matter-matter coupling $G_N$. Note that eqs.~(\ref{Mij})-(\ref{Omega}) correspond to a non-relativistic Kepler orbit, but since $G_N(t)$ evolves in time, the energy and angular momentum are not actually conserved, so we will allow for $\Omega$ to vary, as well as the semi-major axis according to $\dot{a}/a=\dot{G}_N/G_N$ \cite{BisnovatyiKogan:2005ap}. The other parameters in eqs.~(\ref{Mij})-(\ref{Omega}) such as the masses or eccentricity are assumed to be constants.

We then proceed to calculate each term in eq.~(\ref{Pgw1}), accounting for the effects of a time variation of $G_N$ on the binary's orbit. 
The resulting calculations give a number of linear and quadratic derivative terms of couplings, such as $\dot{G}_N$, $\dot{G}_{gw}$, $\dot{G}_{gw}\times \dot{G}_N$, $\dot{G}_N^2$, $\dot{G}_{gw}^2$, and $\Ddot{G}_N$. After taking the time average of these terms, we find that any term that is linear in a single time derivative of either $G_{gw}$ or $G_N$ will vanish leaving only terms that involve quadratic coupling terms or linear terms with higher time derivatives. Whereas contributions from the linear derivative $\dot{G}_N$ vanished in the time average, as we will soon see, there will still be a linear contribution in the total period decay due to direct effects of a varying $G_N$ on orbital motion. Since LLR data already give the tight constraint $|\dot{G}_N/G_N|\lesssim \mathcal{O}(10^{-3})H_0$ and $|\ddot{G}_N/G_N|\lesssim \mathcal{O}(10^{5})H_0^2$, we thus neglect the quadratic term $\dot{G}_N^2$ and the quadratic derivative $\ddot{G}_N$, as both of these terms are expected to be subdominant compared to this linear term $\dot{G}_N$. However, we keep all terms involving $\dot{G}_{gw}$ as we remain agnostic regarding the order of magnitude of this quantity. 

Finally, after taking the time average in $\psi$ and given the assumptions laid out in the previous paragraph, we find that only two non-trivial terms survive in the GW power:
\begin{align} \label{power}
    &P_{gw} =\left[ 1 + K_1\left(\frac{\Dot{G}_N}{G_N}\right)\left(\frac{\Dot{G}_{gw}}{G_{gw}}\right)T_b^{2} + K_2\left(\frac{\Dot{G}_{gw}}{G_{gw}}\right)^2T_b^{2}\right]{P_{GR}}.
\end{align}
Here, the first term corresponds to the standard expression of power radiated in GR generalizing $G$ to $G_{gw}(t)$ (see e.g.~\cite{Beltr_n_Jim_nez_2016}), and comes from $\langle \dddot{Q}_{ij} \dddot{Q}_{ij} \rangle$ in eq.~(\ref{Pgw1}). The second term comes from $\langle \ddot{Q}_{ij} \dddot{Q}_{ij} \rangle$, and the third term from $\langle \ddot{Q}_{ij} \ddot{Q}_{ij} \rangle$. These last two terms describe new contributions that only appear when gravitational couplings have time variations. In eq.~(\ref{power})  we have expressed these contributions in terms of $P_{GR}$ and some dimensionless coefficients $K_{1,2}$ which are functions of the orbital eccentricity and are expected to be order 1 or smaller.

Next, we study the effect this GW emission will have on the period of the binary. In GR, the period is related to the energy as $\Dot{T}_b/T_b =-(3/2)(\Dot{E}/E)$ \cite{Maggiore:1900zz}. We can therefore find the decay of the period by using $\Dot{E} = -P_{gw}$, and expressing the energy $E$ in terms of the period as well as orbital parameters, assuming Newtonian non-relativistic mechanics for the orbit:
\begin{align}\label{dotTb1}
 \dot{T}_{b,GW} = \dot{T}_{b,GR}\left[1 +  K_1\left(\frac{\Dot{G}_N}{G_N}\right)\left(\frac{\Dot{G}_{gw}}{G_{gw}}\right)T_{b}^{2} + K_2\left(\frac{\Dot{G}_{gw}}{G_{gw}}\right)^2T_{b}^{2}
    \right],
\end{align}
where $\dot{T}_{b,GR}$ is the change in the period due to the power $P_{GR}$, which is explicitly given by \cite{Beltr_n_Jim_nez_2016}:
\begin{equation}\label{dotTGR}
    \Dot{T}_{b,GR} = -\frac{192\pi}{5}\left(\frac{G_{gw}}{G_{N}}\right)\frac{G_{N}^{5/3}\mu m^{2/3}}{c^5}\left(\frac{T_b}{2\pi}\right)^{-5/3}f(e).
\end{equation}
Here, we have defined a function of the eccentricity $f(e)$, whose specific expression can be found in \cite{Maggiore:1900zz}.
Eq.~(\ref{dotTb1}) gives the period decay due to GW emission, but it is still missing a contribution from the change in the period just due to the time variations of $G_N$ (which would be present even if there was no GW emission). The analysis of such an effect has been studied previously in \cite{PhysRevLett.61.1151}, where the authors analyzed the motion of the binary system accounting for relativistic effects considering post-Newtonian corrections. They found that a time-varying $G_N$ leads to an additional contribution on the decay of the orbital period, given by $(\Dot{T}_{b}/T_b)=2(\Dot{G}_N/G_N)$. 
Further explorations of a time-varying $G_N$ also include effects on the structure of stars and the subsequent change in total angular momentum of the system \cite{PhysRevLett.65.953, Stairs:2003eg} which result in a model-dependent correction to $\dot{T}_{b}$ that is neglected here.

Finally, accounting for all the previous effects, we obtain that the total time variation of the period is given by:
\begin{align}
    \Dot{T}_{b} &= \Dot{T}_{b, GW}
    + 2T_b\frac{\Dot{G}_N}{G_N}.
\end{align}
With this expression, we next estimate the impact that a variation of $G_{gw}$ on cosmological scales has on binary pulsars. Before specifying values for all these quantities, we note that eq.~(\ref{dotTGR}) reduces to the result in GR when $G_{gw}=G_N=$constant, but here they are changing in time. This time evolution is expected to be slow compared to the observation time so we can approximate it by Taylor expanding and get $G_N^{5/3}(G_{gw}(t)/G_N(t))\approx G_N(t_0)^{5/3}[1+$ $((2/3)\dot{G}_{N}/G_{N}+ \dot{G}_{gw}/G_{gw})(t-t_0)]$, where we have assumed that
$G_{gw}(t_0)\approx G_N(t_0)$. Here, $t_0$ is today and $(t-t_0)$ is the observing time.

Considering the LLR constrains, we assume that $|\dot{G}_N/G_N|\sim \mathcal{O}(10^{-3})H_0$ as well as the expectation that $|\dot{G}_{gw}/G_{gw}|\sim H_0$, as is the case in some dynamical dark energy models. We can then estimate the contributions from the terms arising from time-varying gravitational couplings.
For the Hulse-Taylor pulsar observations \cite{Weisberg:2010zz} $m_{1,2}\sim 1.4M_\odot$, $e\sim 0.6$, and $T_{b}\sim 0.3$days, we estimate $\dot{T}_{b,GR}\sim 2.5 \times 10^{-12} (1+\mathcal{O}(10^{-10})+\mathcal{O}(10^{-13}))$, where we have used that the observing time is of the order of decades. For convenience we also express the binary period $T_b$ in terms of the Hubble rate by noting  $T_b^{-1} \sim 10^{-5} s^{-1}$ and $H_0 \sim 10^{-18} s^{-1}$, which gives $T_b \sim 10^{-13} H_0^{-1}$. We therefore have that:
\begin{align}\label{TdotEstimate}
    \Dot{T}_{b} &\sim \hat{\Dot{T}}_{b,GR} \left[1+\mathcal{O}(10^{-4})+\mathcal{O}(10^{-10})+\mathcal{O}(10^{-13})\right. \nonumber \\ 
   &\left.+\mathcal{O}(10^{-26})+\mathcal{O}(10^{-29}) \right],
\end{align}
where $\hat{\Dot{T}}_{b,GR}$ is the canonical value we would expect in GR, given by eq.\ (\ref{dotTGR}) with $G_{gw}=G_{N}=$const. 
Here we see that the leading order deviation from GR comes from $\dot{G}_N$ originating in the post-Newtonian analysis of the orbital period. The next two relevant terms come from the time evolution of the ratio $G_N^{5/3}(G_{gw}/G_N)$, and the subsequent two terms come from the additional gravitational wave contributions in eq.~(\ref{dotTb1}). This demonstrates that contributions to the orbital period decay from $\dot{G}_{gw}$ are highly suppressed. Observations from the Hulse-Taylor pulsar give $\dot{T}_b/\hat{\Dot{T}}_{b,GR}=1.0032\pm 0.0035 $ \cite{BisnovatyiKogan:2005ap} (and similar in \cite{Weisberg:2010zz}), and thus this data is only sensitive to deviations of GR that are of order $10^{-3}$. Binary pulsars have a sensitivity to set constraints on the leading term in eq.~(\ref{TdotEstimate}) and hence set bounds on $\dot{G}_N/G_N$, while probing a $G_{gw}$ that evolves in a Hubble timescale would require the uncertainty of these observations to be improved by at least 7 orders of magnitude.
We therefore conclude that while the period decay of pulsars does technically depend on the time variation of $G_{gw}$, observational data does not allow us to place tight constraints on it, and hence standard sirens will become the best probe to set interesting bounds on $G_{gw}$ in the future.


We finally note that other constraints on time-varying gravitational couplings have been placed from primordial abundances of light elements \cite{MALANEY1993145, Bambi:2005fi, Clifton:2005xr}, or  planetary radar-ranging measurements \cite{Turyshev:2004nu, Pitjeva2005, Konopliv2011, Pitjeva:2013fja}, however none of these constraints apply to $G_{gw}$ so they have not been discussed further. 

\textit{Discussion.---}
Dynamical dark energy cosmological models typically introduce modifications to the gravitational sector, and hence can be tested with observations of gravitational waves. In this Letter we focus on the possible presence of a time-varying local gravitational self-interaction coupling $G_{gw}$, whose time evolution can be probed with standard sirens. We first showed that simple modifications to the $\Lambda$CDM model in which dark energy is described by a single scalar field exhibit a different gravitational coupling for matter-matter interactions, $G_N$, and for metric self-interactions $G_{gw}$. Therefore, it is expected that general dark energy models predict $G_N\not=G_{gw}$. We then explored how other local observations may have already probed the time evolution of $G_{gw}$, and hence could be used to shed light on the outlook of standard sirens. We argued that LLR observations constrain the time variation of $G_N$ and thus are in general irrelevant for standard sirens, but binary pulsar observations do indeed probe $G_{gw}$. In particular, we found that while the time variation of both $G_N$ and $G_{gw}$ affect the binary's period decay, the effects from $\Dot{G}_{gw}$ are highly suppressed, and thus pulsars place extremely weak and uninformative constraints on it. We therefore conclude that future standard sirens will provide the tightest constraints on the time variation of $G_{gw}$ and hence they will provide crucial information on the properties of dark energy. 

Finally, we note that the analysis performed here included a number of simplifying assumptions, such as neglecting other effects that may appear in dynamical dark energy models. Additional contributions to the binary's period decay may come from dipole radiation or scalarization (see e.g.~\cite{2009MNRAS.400..805L,Will:2014kxa, Manchester:2015mda, Anderson:2019eay, Renevey:2019jrm}). In order to quantify these effects one must specify the underlying theory considered, however in this Letter we have taken an agnostic approach. Nevertheless, the addition of all these contributions would weaken the potential constraints one may impose on $\dot{G}_{gw}$ with binary pulsars, and hence they do not affect the main conclusions of this analysis. 

\vspace{0.2in}

\section*{Acknowledgments}
\vspace{-0.2in}
We are grateful to discussions with Wayne Hu and Jose Maria Ezquiaga. WW was supported at the University of Chicago by the Kavli Institute for Cosmological Physics. ML was supported at the University of Chicago by the Kavli Institute for Cosmological Physics through an endowment from the Kavli Foundation and its founder Fred Kavli.

\bibliography{RefModifiedGravity} 

 \end{document}